\documentclass[a4paper,12pt]{article}
\usepackage{amsmath}
\usepackage{amssymb}
\usepackage{amsmath,amsthm,amssymb,amscd}
\usepackage{latexsym}
\usepackage{indentfirst}
\usepackage{graphicx}
\usepackage[top=1in,bottom=1in,left=1.25in,right=1.25in]{geometry}

\makeatother
\date{}
\theoremstyle{plain}

\begin{document}

\bibliographystyle{unsrt}
\bibliographystyle{plain}

\title{
  \bf A general two-cycle network model of molecular motors
}

\author{{Yunxin Zhang}
\thanks{School of Mathematical Sciences, Fudan University,  Shanghai 200433,
China  (E-Mail: xyz@fudan.edu.cn)}\thanks{Centre for Computational
Systems Biology, Fudan University }\thanks{Shanghai Key Laboratory
for Contemporary Applied Mathematics, Fudan University} }

%\date{\today}

\maketitle \baselineskip=6mm

\baselineskip=22pt

\begin{abstract}
Molecular motors are single macromolecules that generate forces at
the piconewton range and nanometer scale. They convert chemical
energy into mechanical work by moving along filamentous structures.
In this paper, we study the velocity of two-head molecular motors in the
framework of a mechanochemical network theory.  The network model, a
generalization of the recently work of Liepelt and Lipowsky (PRL 98,
258102 (2007)), is based on the discrete mechanochemical states of a
molecular motor with multiple cycles.  By generalizing the
mathematical method developed by Fisher and Kolomeisky for single
cycle motor (PNAS(2001) 98(14) P7748-7753), we are able to obtain an
explicit formula for the velocity of a molecular motor.
% with nonzero
%ATP hydrolysis rate even under the stalling condition.
%The general
%multi-cycle network model is an important intermediate between the
%ratchet models and the discrete, single-cycle chemical models.

\vspace{2em} \noindent \textit{PACS}: 87.16.Nn, 87.16.A-, 82.39.-k,
05.40.Jc

\vspace{2em} \noindent \textit{Keywords}: molecular motors,
mechanochemical network
\end{abstract}

\section{Introduction}
In biological cells, molecular motors are individual protein
molecules that are responsible for many of the biophysical
functions of the cellular movement and mechanics.
Important examples of motor proteins are kinesin
\cite{David2007,Carter2005,Taniguchi2005}, dynein
\cite{Vale2003,Sakakibara1999}, mysion \cite{Hooft2007,Christof2006,
Katsuyuki2007} and $F_0F_1$-ATP synthase \cite{Noji1997}. Molecular
motors are mechanochemical force generators which convert
biochemical energy (stored as ATP, adenosine triphosphate) into
mechanical work in a thermal environment \cite{Howard2001, Zhang2009}.
Many molecular motors, due to their two-head nature and hand-over-hand
mechanism, can move processively along their tracks for a
long time before its dissociation from the track. For example, myosin
slides along an actin filament, kinesin and dynein along microtubule
(MT). The velocity of molecular motors is quite fast, with mean
velocity at about several hundreds nanometers per second \cite{voboda1994}.
Understanding how the various molecular motors operate is a
significant scientific challenge with important nano-engineering implications.

To understand the principle of molecular motors, a good mathematical
model is essential.  Much progress has been made in recent years in
theoretical analysis of molecular motors. Mainly two different
approaches have been taken: The ratchet models that consider motor
chemical transitions occur without explicit coupling to motor
steppings \cite{Prost1997,Qian2005}, and the discrete chemical
models that contain only a single chemomechanical cycle
\cite{Kolomeisky2007,Qian1997}. Recently, however, Liepelt and
Lipowsky \cite{Liepelt2007,Liepelt2009} introduced a six-state
network to model the chemomechanical motor cycles, in which the
dynamics of two-head motor molecule is described by a Markovian jump
process. In \cite{Schmiedl2008}, Schmiedl
 and Seifert used
a two states network to discuss the efficiency of the molecular
motors. The importance of the latter development is in introducing
futile cycles into the discrete chemical model, thus making the
discrete chemical approach and continuous Brownian approach more
connected. %One important prediction of the latter models is the
%non-zero hydrolysis under the stalling condition.
Their results indicate that the network modeling approach is a good
choice for the theoretical analysis of the molecular motors.  In the
past, a great deal of mathematical analysis is based on the Brownian
ratchet formalism. Similar network models has also be used
successfully in the theoretical analysis of other biochemical
processes \cite{Cady2009,Lau2007}.

In this paper, we shall generalize the network model to include
arbitrary $2N$ number of states. In particular, we shall use the
network model to analyze the movement of molecular motors.
Mathematically, therefore, the models developed in
\cite{Liepelt2009, Schmiedl2008} and even those in \cite{Cady2009,
Lau2007} can be regarded as special cases of our network model. In
the framework of this network model, we further develop a method
pioneered by Derrida, Fisher and Kolomeisky \cite{Fisher2001,
Kolomeisky2000,Derrida1983} to compute the mean velocity of a
molecular motor.

\begin{figure}
  \centering
  \includegraphics[width=400pt]{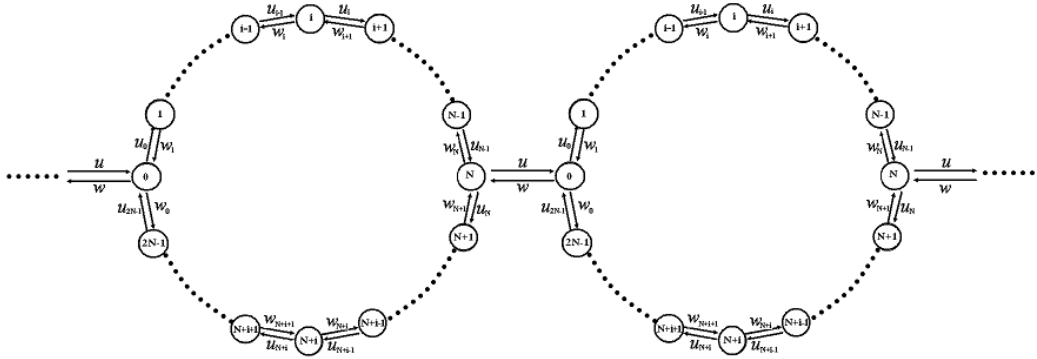}\\
  \caption{A schematic depiction of the $2N$ states network model for molecular motors. One forward step of molecular motors is completed only
  in the biochemical process $N\to 0$. In mechanochemical process $0\to 1\to\cdots\to N\to
N+1\to\cdots\to 2N-1\to 0$, the molecular motors make no mechanical
step while hydrolyzing two ATP.}\label{Figure1}
\end{figure}

In our model, a two-head molecular motor with hand-over-hand
mechanism is assumed to have $2N$
mechanochemical states in their movement, denoted by $0, 1, 2,
\cdots, 2N-1$ (see Figure \ref{Figure1}).   The two heads moves
exactly with half of cycle out of phase.  If there are 2N states
in the hydrolysis kinetic cycle of a single head; we have states
$(0,N), (1,N+1), (2,N+2), \cdots, (2N-2,N-2)$, and $(2N-1,N-1)$
for the motor with two heads.  The
hand-over-hand mechanism means the motor ``walks'' a step with
the transition $(N,0)\rightarrow (0,N)$, switching the leading
and the trailing head.  However, it is possible that the
translocation does not occur, and the kinetic cycle is
completed as a futile cycle, with two ATP hydrolyzed, one by
each head.

    From now one, we shall use the state of the leading head to
denote the state of the motor; and denote the forward and
backward rate parameters at state $i$ as $u_i$ (i.e., $i\to i+1$) and
$w_i$ ($i\to i-1$) respectively, which satisfy $u_{2N+i}=u_i$ and
$w_{2N+i}=w_i$ (since molecular motors move forward periodically).
Generally, the transition rates $u_i$ and $w_i$ depend on the
external force $F_{ext}$ and the free energy $\triangle G$ released
by the fuel molecular. The transition rates between state $N$ and
$0$, the hand-over-hand, are $u$ and $w$.   In the following, we
suppose that all these transition rates are known explicitly.

    The transition from $N\to 0$ represents the switching between the
leading and trailing heads, thus moves one motor step.
If a mechanochemical process takes $0\to 1\to\cdots\to N\to
N+1\to\cdots\to 2N-1\to 0$, the molecular motors make no
mechanical step while hydrolyzing two ATP.  However, if the
process takes $0\to 1\to\cdots\to N\to 0\to 1\to\cdots N\to 0$,
then the motor hydrolyzed two ATP and moved two steps.  It can
be readily found that, for $N=3$, this model reduces to the 6 states
network model in \cite{Liepelt2007}, for $N=1$, this model reduces
to the 2 states model in \cite{Schmiedl2008}.

    In the next section,
we shall give the formulation of the velocity of molecular motors
using the network model. We will discuss some special cases in
section 3.  The force dependence of the transition rates $u_i, w_i$
and $u, w$ is discussed in section 4. In section 5, we will discuss
the continuous mechanochemical sate case of our multi-cycle model,
and section 6 contains concluding remarks.

\section{ The velocity of molecular motors}
In this section, we will calculate the velocity of the molecular
motors in the framework of our network model. The method used in the
following is similar to the one used in \cite{Fisher2001,
Kolomeisky2000} and \cite{Derrida1983}.

Let $\rho_{i}(t)$ be the probability density for finding molecular
motors in
state $i$ at time $t$. %$u_i$ and $w_i$ be the forward and backward rate
%parameters.
The evolution of the probability density $\rho_i(t)$ is governed by
the following master equations
\begin{equation}\label{eq1}
\begin{aligned}
\frac{d\rho_i}{dt}=&(\rho_{i-1}u_{i-1}+\rho_{i+1}w_{i+1})-\rho_{i}(u_{i}+w_{i})\cr
=&(\rho_{i-1}u_{i-1}-\rho_{i}w_{i})-(\rho_{i}u_{i}-\rho_{i+1}w_{i+1})\cr
\triangleq&J_i-J_{i+1}\qquad \textrm{for}\ 1\le i\le N-1\ \textrm{or}\
N+1\le i\le 2N-1
\end{aligned}
\end{equation}
and
\begin{equation}\label{eq2}
\begin{aligned}
&\begin{aligned}
\frac{d\rho_0}{dt}=&(\rho_{2N-1}u_{2N-1}-\rho_{0}w_{0})-(\rho_{0}u_{0}-\rho_{1}w_{1})+(\rho_{N}u-\rho_0w)\cr
\triangleq&J_{2N}-J_{1}+J
\end{aligned}\cr
&\begin{aligned}
\frac{d\rho_N}{dt}=&(\rho_{N-1}u_{N-1}-\rho_{N}w_{N})-(\rho_{N}u_{N}-\rho_{N+1}w_{N+1})-(\rho_{N}u-\rho_0w)\cr
\triangleq&J_{N}-J_{N+1}-J
\end{aligned}
\end{aligned}
\end{equation}
where
\begin{equation}\label{eq3}
J_i=\rho_{i-1}u_{i-1}-\rho_{i}w_{i}\qquad J=\rho_{N}u-\rho_0w
\end{equation}
$J_i$ is the probability flux from mechanochemical state $i-1$ to
state $i$, and $J$ is the probability flux from mechanochemical
state $N$ to state $0$. At steady state,
\begin{equation}\label{eq4}
J_1=J_2=\cdots=J_N\quad J_{N+1}=J_{N+2}=\cdots=J_{2N}\quad
J_1=J_{2N}+J
\end{equation}
By Eqs. (\ref{eq1}-\ref{eq4}), one can know that
\begin{equation}\label{eq5}
\begin{aligned}
&\rho_k=\rho_0\prod_{i=1}^k\left(\frac{u_{i-1}}{w_i}\right)-\left[1+\sum_{i=1}^{k-1}\prod_{j=i}^{k-1}\left(\frac{u_{j}}{w_j}\right)\right]\frac{J_1}{w_k}\cr
&\rho_{N+k}=\rho_N\prod_{i=1}^k\left(\frac{u_{N+i-1}}{w_{N+i}}\right)-\left[1+\sum_{i=1}^{k-1}\prod_{j=i}^{k-1}\left(\frac{u_{N+j}}{w_{N+j}}\right)\right]\frac{J_1-J}{w_{N+k}}
\end{aligned}
\end{equation}
where $1\le k\le N-1$ and
\begin{equation}\label{eq6}
\begin{aligned}
&\rho_N=\rho_0\prod_{i=1}^N\left(\frac{u_{i-1}}{w_i}\right)-\left[1+\sum_{i=1}^{N-1}\prod_{j=i}^{N-1}\left(\frac{u_{j}}{w_j}\right)\right]\frac{J_1}{w_N}
\end{aligned}
\end{equation}
$J_{2N}=J_1-J$ means
$\rho_{2N-1}u_{2N-1}-\rho_{0}w_{0}=J_1-J$, %(\rho_{N}u-\rho_0w)$, i.e.
so %\rho_{N}u+\rho_{2N-1}u_{2N-1}-\rho_{0}(w+w_{0})=J_1
\begin{equation}\label{eq7}
\begin{aligned}
\rho_0=&\frac{\rho_{2N-1}u_{2N-1}}{w_{0}}-\frac{J_1-J}{w_{0}}\cr
%=&\frac{u_{2N-1}}{w_{0}}\left\{\rho_N\prod_{i=1}^{N-1}\left(\frac{u_{N+i-1}}{w_{N+i}}\right)-\left[1+\sum_{i=1}^{N-2}\prod_{j=i}^{N-2}\left(\frac{u_{N+j}}{w_{N+j}}\right)\right]\frac{J_1-J}{w_{2N-1}}\right\}-\frac{J_1-J}{w_{0}}\cr
=&\rho_N\prod_{i=1}^{N}\left(\frac{u_{N+i-1}}{w_{N+i}}\right)-\left[1+\sum_{i=1}^{N-1}\prod_{j=i}^{N-1}\left(\frac{u_{N+j}}{w_{N+j}}\right)\right]\frac{J_1-(\rho_{N}u-\rho_0w)}{w_{0}}\cr
\end{aligned}
\end{equation}
Substituting (\ref{eq6}) into (\ref{eq7}), one obtains
\begin{equation}\label{eq9}
\rho_0=\frac{A}{B}J_1
\end{equation}
where
\begin{equation}\label{eq10}
\begin{aligned}
A=\left[1+\sum_{i=1}^{2N-1}\prod_{j=i}^{2N-1}\left(\frac{u_{j}}{w_{j}}\right)\right]\frac{1}{w_{0}}+\left[1+\sum_{i=1}^{N-1}\prod_{j=i}^{N-1}\left(\frac{u_{j}}{w_j}\right)\right]\left[1+\sum_{i=N+1}^{2N-1}\prod_{j=i}^{2N-1}\left(\frac{u_{j}}{w_{j}}\right)\right]\frac{u}{w_{0}w_N}
\end{aligned}
\end{equation}
and
\begin{equation}\label{eq11}
\begin{aligned}
B=\left[\frac{u}{w_{N}}\prod_{i=0}^{N-1}\left(\frac{u_{i}}{w_i}\right)-\frac{w}{w_{0}}\right]\left[1+\sum_{i=N+1}^{2N-1}\prod_{j=i}^{2N-1}\left(\frac{u_{j}}{w_{j}}\right)\right]+\prod_{i=0}^{2N-1}\left(\frac{u_{i}}{w_i}\right)-1
\end{aligned}
\end{equation}
So
\begin{equation}\label{eq12}
\begin{aligned}
J=\left\{\left[u\prod_{i=1}^N\left(\frac{u_{i-1}}{w_i}\right)-w\right]\frac{A}{B}-\left[1+\sum_{i=1}^{N-1}\prod_{j=i}^{N-1}\left(\frac{u_{j}}{w_j}\right)\right]\frac{u}{w_N}\right\}J_1%\cr
=:CJ_1
\end{aligned}
\end{equation}
By (\ref{eq5}) (\ref{eq6}) (\ref{eq9}) (\ref{eq12}), we get the
expressions of probabities $\rho_k$ and $\rho_{N+k}$ as functions of
$J_1$:
\begin{equation}\label{eq13}
\begin{aligned}
\rho_k%=&\left\{\frac{A}{B}\prod_{i=1}^k\left(\frac{u_{i-1}}{w_i}\right)-\left[1+\sum_{i=1}^{k-1}\prod_{j=i}^{k-1}\left(\frac{u_{j}}{w_j}\right)\right]\frac{1}{w_k}\right\}J_1\cr
=\left\{\frac{Aw_0}{B}\prod_{i=0}^{k-1}\left(\frac{u_{i}}{w_i}\right)-\sum_{i=1}^{k-1}\prod_{j=i}^{k-1}\left(\frac{u_{j}}{w_j}\right)-1\right\}\frac{J_1}{w_k}
\end{aligned}
\end{equation}
\begin{equation}\label{eq14}
\begin{aligned}
\rho_{N+k}=&\left\{\frac{Aw_0}{B}\prod_{i=0}^{N+k-1}\left(\frac{u_{i}}{w_i}\right)-\sum_{i=1}^{N+k-1}\prod_{j=i}^{N+k-1}\left(\frac{u_{j}}{w_j}\right)\right.\cr
&+\left.C\sum_{i=N+1}^{N+k-1}\prod_{j=i}^{N+k-1}\left(\frac{u_{j}}{w_{j}}\right)-(1-C)\right\}\frac{J_1}{w_{N+k}}\cr
\end{aligned}
\end{equation}
Conservation of probability requires
\begin{equation}\label{eq15}
\sum_{k=0}^{2N-1}\rho_{k}=1
\end{equation}
So, from (\ref{eq13}) (\ref{eq14}) (\ref{eq15}), one knows
\begin{equation}\label{eq16}
J_1=\frac{1}{D}
\end{equation}
where
\begin{equation}\label{eq17}
\begin{aligned}
D=&\frac{A}{B}\left\{\sum_{k=1}^{2N-1}\left[\frac{w_0}{w_k}\prod_{i=0}^{k-1}\left(\frac{u_{i}}{w_i}\right)\right]+1\right\}-\sum_{k=1}^{2N-1}\left[\frac{1}{w_k}\sum_{i=1}^{k-1}\prod_{j=i}^{k-1}\left(\frac{u_{j}}{w_j}\right)\right]\cr
&-\sum_{k=1}^{2N-1}\left(\frac{1}{w_k}\right)+C\sum_{k=N+1}^{2N-1}\frac{1}{w_{k}}\left[1+\sum_{i=N+1}^{k-1}\prod_{j=i}^{k-1}\left(\frac{u_{j}}{w_{j}}\right)\right]
\end{aligned}
\end{equation}
In view of (\ref{eq12}) and (\ref{eq16}), one obtains
\begin{equation}\label{eq18}
\begin{aligned}
J=CJ_1=\frac{C}{D}
\end{aligned}
\end{equation}
So the mean velocity of the molecular motors is
\begin{equation}\label{eq19}
\begin{aligned}
V=JL=\frac{CL}{D}
\end{aligned}
\end{equation}
where $L$ is the stepsize of the molecular motors (8.2nm for motor
protein kinesin). Certainly, the explicit expresions of
probabilities $\rho_k (0\le k\le 2N-1)$ also can be obtained by
(\ref{eq6}) (\ref{eq9}) (\ref{eq13}) (\ref{eq14}).

\section{The special cases of the network model}
In this section, we consider some special cases of the network
model.

\noindent{\bf (1)} $w_0=w_{2N-1}=w_{2N-2}=\cdots=w_{N+1}=0$, and
$u_N=u_{N+1}=\cdots=u_{2N-1}=0$ (see Figure \ref{Figure2} {\bf
(Up)}):
\begin{figure}
  \includegraphics[width=400pt]{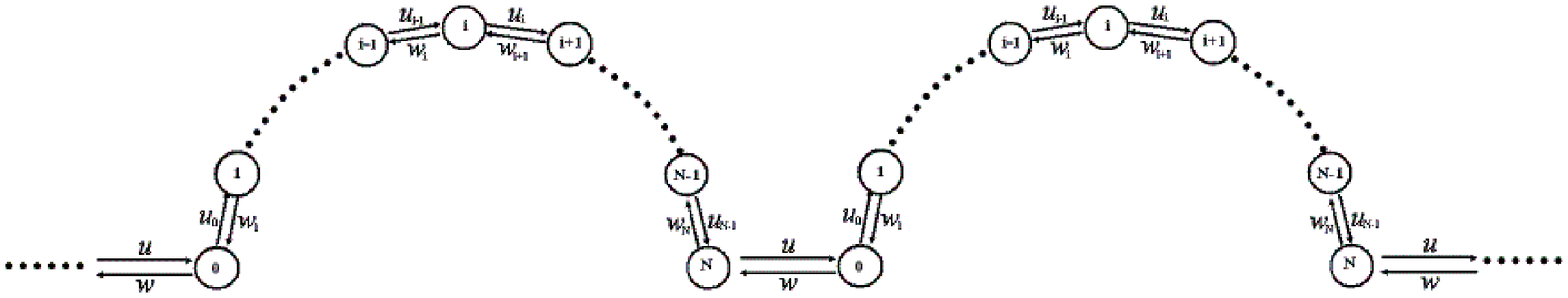}\\\includegraphics[width=400pt]{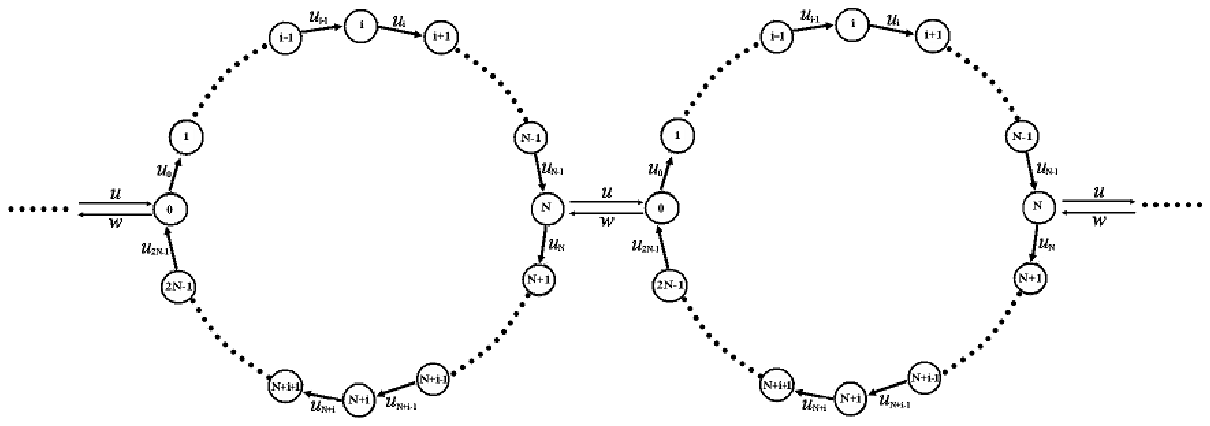}\\
  \caption{Special cases of the network model: {\bf (Up)}
   in which $w_0=w_{2N-1}=w_{2N-2}=\cdots=w_{N+1}=0$, and
$u_N=u_{N+1}=\cdots=u_{2N-1}=0$. {\bf (Down)} in which
$w_0=w_1=\cdots=w_{2N-1}=0$. }\label{Figure2}
\end{figure}

In this case, our network model reduces to the usual one dimensional
hopping model \cite{Derrida1983, Kolomeisky2000}. It can be easily
found that $J_i=\rho_{i-1}u_{i-1}-\rho_{i}w_{i}=0$ for $N+1\le i\le
2N$, and $J=\rho_Nu-\rho_0w=\rho_{0}u_0-\rho_1w_1=J_1$. By
$J_1=J_2=\cdots=J_N$, one obtains
\begin{equation}\label{eq20}
\begin{aligned}
&\rho_k=\rho_0\prod_{i=1}^k\left(\frac{u_{i-1}}{w_i}\right)-\left[1+\sum_{i=1}^{k-1}\prod_{j=i}^{k-1}\left(\frac{u_{j}}{w_j}\right)\right]\frac{J}{w_k}\cr
\end{aligned}
\end{equation}
and
$\rho_N=\rho_0\prod_{i=1}^N\left(\frac{u_{i-1}}{w_i}\right)-\left[1+\sum_{i=1}^{N-1}\prod_{j=i}^{N-1}\left(\frac{u_{j}}{w_j}\right)\right]\frac{J}{w_N}$.
At the same time, $\rho_Nu-\rho_0w=J$ implies
\begin{equation}\label{eq21}
\begin{aligned}
\frac{J+\rho_0w}{u}=\rho_N=\rho_0\prod_{i=1}^N\left(\frac{u_{i-1}}{w_i}\right)-\left[1+\sum_{i=1}^{N-1}\prod_{j=i}^{N-1}\left(\frac{u_{j}}{w_j}\right)\right]\frac{J}{w_N}
\end{aligned}
\end{equation}
which gives
\begin{equation}\label{eq22}
\begin{aligned}
\rho_0=\frac{1+\left[1+\sum_{i=1}^{N-1}\prod_{j=i}^{N-1}\left(\frac{u_{j}}{w_j}\right)\right]\frac{u}{w_N}}{u\prod_{i=1}^N\left(\frac{u_{i-1}}{w_i}\right)-w}J
\end{aligned}
\end{equation}
Combing (\ref{eq20}) (\ref{eq22}), we get
\begin{equation}\label{eq23}
\begin{aligned}
&\rho_k=\left\{\frac{1+\left[1+\sum_{i=1}^{N-1}\prod_{j=i}^{N-1}\left(\frac{u_{j}}{w_j}\right)\right]\frac{u}{w_N}}{u\prod_{i=1}^N\left(\frac{u_{i-1}}{w_i}\right)-w}\prod_{i=1}^k\left(\frac{u_{i-1}}{w_i}\right)-\left[1+\sum_{i=1}^{k-1}\prod_{j=i}^{k-1}\left(\frac{u_{j}}{w_j}\right)\right]\frac{1}{w_k}\right\}J\cr
\end{aligned}
\end{equation}
Finally, $\sum_{k=0}^{N}\rho_{k}=1$ gives
\begin{equation}\label{eq24}
\begin{aligned}
J=\frac{1}{\bar A}
\end{aligned}
\end{equation}
where
\begin{equation}\label{eq25}
\begin{aligned}
\bar
A=\frac{1+\left[1+\sum_{i=1}^{N-1}\prod_{j=i}^{N-1}\left(\frac{u_{j}}{w_j}\right)\right]\frac{u}{w_N}}{u\prod_{i=1}^N\left(\frac{u_{i-1}}{w_i}\right)-w}\sum_{k=0}^{N}\left[\prod_{i=1}^k\left(\frac{u_{i-1}}{w_i}\right)\right]-\sum_{k=0}^{N}\left\{\left[1+\sum_{i=1}^{k-1}\prod_{j=i}^{k-1}\left(\frac{u_{j}}{w_j}\right)\right]\frac{1}{w_k}\right\}
\end{aligned}
\end{equation}
So in this case, the mean velocity of molecular motors is
$V=JL=L/\bar A$, and the probabilities $\rho_k$ are given by Eqs.
(\ref{eq22}) (\ref{eq23}).

\noindent{\bf (2)} $w_0=w_1=\cdots=w_{2N-1}=0$ (see Figure
\ref{Figure2}  {\bf (Down)}):

In this case, $J_i=\rho_{i-1}u_{i-1}$ for $1\le i\le 2N$, and
$J=\rho_Nu-\rho_0w$. At the steady state
\begin{equation}\label{eq26}
\begin{aligned}
\rho_k=\frac{u_0}{u_k}\rho_0\qquad
\rho_{N+k}=\frac{u_N}{u_{N+k}}\rho_N\quad \textrm{for}\quad 0\le
k\le N-1
\end{aligned}
\end{equation}
Due to $J_{N+1}+J=J+{N-1}$, one knows
\begin{equation}\label{eq27}
\begin{aligned}
\rho_N(u+u_N)=\rho_0(u_0+w)
\end{aligned}
\end{equation}
i.e.
\begin{equation}\label{eq28}
\begin{aligned}
\rho_N=\frac{u_0+w}{u+u_N}\rho_0
\end{aligned}
\end{equation}
in view of (\ref{eq26}) (\ref{eq28}) and
$\sum_{k=0}^{2N-1}\rho_{k}=1$, one obtains
\begin{equation}\label{eq29}
\begin{aligned}
\rho_0=\frac{1}{u_0\sum_{k=0}^{N-1}\frac{1}{u_k}+\frac{u_N(u_0+w)}{u+u_N}\sum_{k=N}^{2N-1}\frac{1}{u_k}}
\end{aligned}
\end{equation}
hence
\begin{equation}\label{eq30}
\begin{aligned}
J=&\rho_Nu-\rho_0w=\left(\frac{u(u_0+w)}{u+u_N}-w\right)\rho_0\cr
=&\frac{uu_0-u_Nw}{u_0(u+u_N)\sum_{k=0}^{N-1}\frac{1}{u_k}+u_N(u_0+w)\sum_{k=N}^{2N-1}\frac{1}{u_k}}
\end{aligned}
\end{equation}
and the probabilities $\rho_k$ can be obtained by Eqs. (\ref{eq26})
(\ref{eq28}) (\ref{eq29}).

\noindent{\bf (3)} $N=1$:

In this case,
\begin{equation}\label{eq31}
\begin{aligned}
A=\frac{u_{{1}}+u+w_{{1}}}{w_0w_{{1}}}\qquad
B=\frac{uu_{{0}}-ww_{{1}}+u_{{0}}u_{{1}}-w_{{1}}w_{{0}}}{w_{{1}}w_{{0}}}
\end{aligned}
\end{equation}
$$
C=\left( {\frac {uu_{{0}}}{w_{{1}}}}-w \right)  \left(
{\frac{u_{{1}}} {w_{{1}}w_{{0}}}}+{w_{{0}}}^{-1}+{\frac
{u}{w_{{1}}w_{{0}}}} \right)
 \left( {\frac {uu_{{0}}}{w_{{1}}w_{{0}}}}-{\frac {w}{w_{{0}}}}+{
\frac {u_{{0}}u_{{1}}}{w_{{1}}w_{{0}}}}-1
\right)^{-1}-{\frac{u}{w_{ {N}}}}
$$
$$
D=\left( {\frac {u_{{1}}}{w_{{1}}w_{{0}}}}+{w_{{0}}}^{-1}+{\frac
{u}{w_ {{1}}w_{{0}}}} \right)  \left( {\frac {u_{{0}}}{w_{{1}}}}+1
\right)\left( {\frac
{uu_{{0}}}{w_{{1}}w_{{0}}}}-{\frac{w}{w_{{0}}}}+{ \frac
{u_{{0}}u_{{1}}}{w_{{1}}w_{{0}}}}-1 \right) ^{-1}-{w_{{1}}}^{-1}
$$
and
\begin{equation}\label{eq32}
\begin{aligned}
J=\frac{\frac{\left( {\frac {uu_{{0}}}{w_{{1}}}}-w \right)  \left(
{\frac { u_{{1}}}{w_{{1}}w_{{0}}}}+{w_{{0}}}^{-1}+{\frac
{u}{w_{{1}}w_{{0}}}}
 \right)}{ {\frac {uu_{{0}}}{w_{{1}}w_{{0}}}}-{\frac {w}{w_{{0}}
}}+{\frac {u_{{0}}u_{{1}}}{w_{{1}}w_{{0}}}}-1}-{\frac {u}
{w_{{N}}}}}{ \frac{\left( {\frac{u_{{1}}}{w_{{1}}w_{{0}}}}+{
w_{{0}}}^{-1}+{\frac {u}{w_{{1}}w_{{0}}}} \right)  \left( {\frac
{u_{{0 }}}{w_{{1}}}}+1 \right)}{{\frac
{uu_{{0}}}{w_{{1}}w_{{0}}}}-{ \frac {w}{w_{{0}}}}+{\frac
{u_{{0}}u_{{1}}}{w_{{1}}w_{{0}}}}-1}-\frac{1}{w_1}}
\end{aligned}
\end{equation}
The probability flux of the special case (1) is
\begin{equation}\label{eq33}
\begin{aligned}
J={\frac {uu_{{0}}-ww_{{1}}}{u_{{0}}+w_{{1}}+u+w}}
\end{aligned}
\end{equation}
The probability flux of the special case (2) is
\begin{equation}\label{eq34}
\begin{aligned}
J={\frac {uu_{{0}}-u_{{1}}w}{u+u_{{1}}+u_{{0}}+w}}
\end{aligned}
\end{equation}

\noindent{\bf (4)} $N=2$:

In this case, the probability flux (\ref{eq24}) is
\begin{equation}\label{eq35}
\begin{aligned}
J={\frac
{w_{{1}}w_{{2}}w_{{3}}w_{{0}}}{u_{{1}}u_{{2}}u_{{3}}+u_{{2}}u_{{3}}w_{{1}}+w_{{1}}w_{{2}}w_{{3}}+w_{{1}}w_{{2}}u_{{3}}+uw_{{1}}w_{{3}
}+uw_{{1}}u_{{3}}+uu_{{1}}w_{{3}}+uu_{{1}}u_{{3}}}}
\end{aligned}
\end{equation}
the probability flux (\ref{eq30}) is
\begin{equation}\label{eq36}
\begin{aligned}
J={\frac { \left( uu_{{0}}-u_{{2}}w
\right)u_{{1}}u_{{3}}}{uu_{{1}}u_{{
3}}+u_{{3}}uu_{{0}}+u_{{1}}u_{{2}}u_{{3}}+u_{{3}}u_{{0}}u_{{2}}+u_{{1}
}u_{{0}}u_{{3}}+u_{{1}}u_{{0}}u_{{2}}+u_{{1}}wu_{{3}}+u_{{1}}u_{{2}}w}}
\end{aligned}
\end{equation}

\section{The force dependence of the transition rates}

\begin{figure}
  \includegraphics[width=200pt]{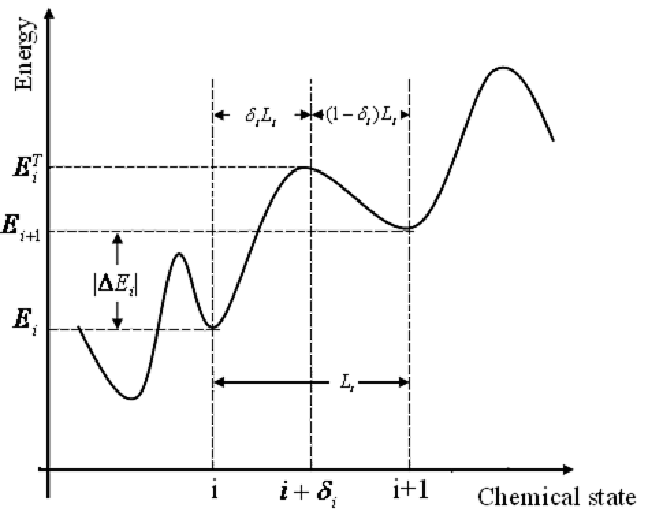}\includegraphics[width=200pt]{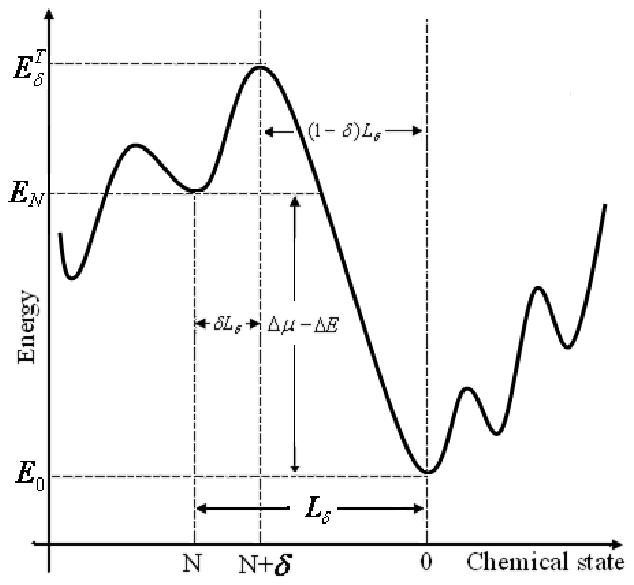}\\
  \caption{Energy profile of a  molecular motor in the neighborhood of local equilibrium mechanochemical state: %({\bf Left}) $\delta_i, L_i$  and $\delta, L$ ({\bf Right})
  ({\bf Left}) Molecular motor undergoes thermal fluctuations around the $i-$th local equilibrium position with potential
  $E_i$, which corresponds to mechanochemical state $i$. It moves forward (to the right) or backward (to the left) when it
  acquires enough energy to across the energy barriers $E^T_i$ or $E^T_{i-1}$. The local equilibrium position $i$ and $i+1$ are separated by characteristic distance $L_i$,
  the local equilibrium state $i$ and the transition state ${i+\delta_i}$ with energy $E^T_i$ are separated by characteristic distance
  $\delta_iL_i$, and the local equilibrium state $i+1$ and the transition state ${i+\delta_i}$ are separated by characteristic distance
  $(1-\delta_i)L_i$. The energy difference between state $i$ and
  $i+1$ is $\triangle E_i=E_i-E_{i+1}$.
  ({\bf Right}) Molecular motor undergoes thermal fluctuations around the $N-$th local equilibrium position with potential
  $E_N$, which corresponds to mechanochemical state $N$. It moves forward (to the right) when it
  acquires enough energy to across the energy barriers $E^T_\delta$.
  The local equilibrium $N$ and $0$ are separated by characteristic distance $L_\delta$,
  the local equilibrium $N$ and the transition state $N+\delta$ with energy $E^T_{\delta}$ are separated by characteristic distance $\delta L_\delta$,
  the local equilibrium $0$ and the transition state $N+\delta$ are separated by characteristic distance $(1-\delta)L_\delta$. The energy difference
  between state $0$ and $N$ is $\triangle\mu-\triangle E$.}
  \label{Figure3}
\end{figure}

As pointed out in the introduction, the transition rates $u_i, w_i,
u, w$ depend on the external force $F$. For nonzero external force
$F$, the force dependence of the transition rates $u_i, w_i, u, w$
can be modeled as the following
\begin{equation}\label{eq37}
\begin{aligned}
&u=k^+e^{-\beta\delta FL_{\delta}}\ \ \quad w=k^-e^{\beta(1-\delta)
FL_{\delta}}\cr &u_i=k_i^+e^{-\beta\delta_i FL_i}\quad
w_{i+1}=k_{i+1}^-e^{\beta(1-\delta_i) FL_i}\qquad 0\le i\le 2N-1
\end{aligned}
\end{equation}
where $\beta=1/k_BT$, $0\le \delta, \delta_i\le 1$ are load
distribution factors that reflect how the external force affects the
individual rates \cite{Schmiedl2008, Fisher2001, Lau2007} (see
Figure \ref{Figure3}), $L_0+L_1+\cdots+L_{N-1}+L_{\delta}=L$,
$L_0+L_1+\cdots+L_{N-1}=L_N+L_{N+1}+\cdots+L_{2N-1}$.

\begin{figure}
  \center
  \includegraphics[width=200pt]{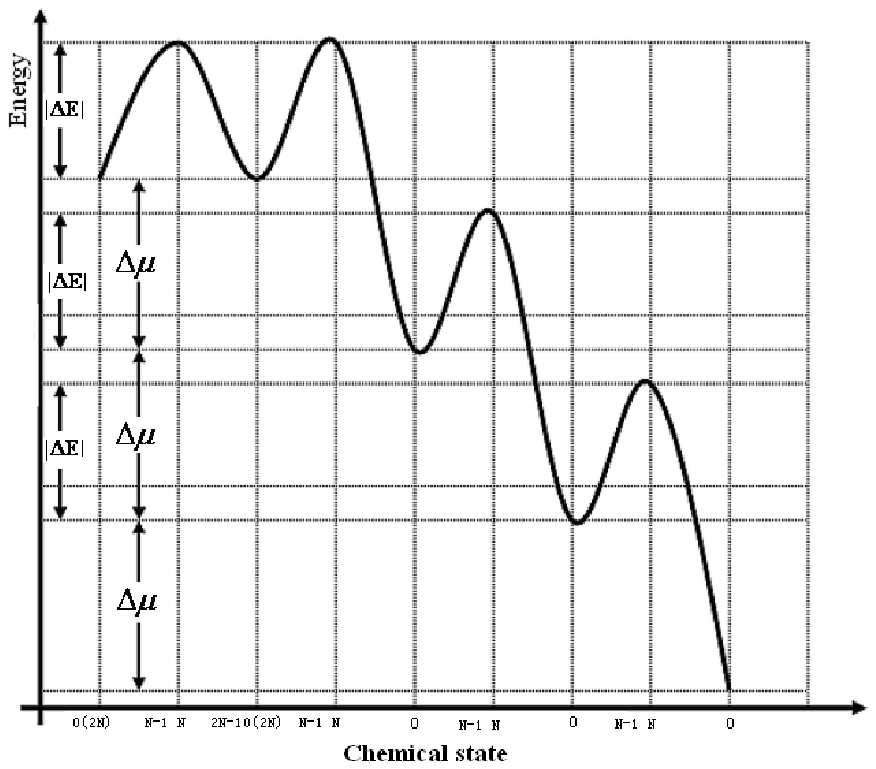}\includegraphics[width=200pt]{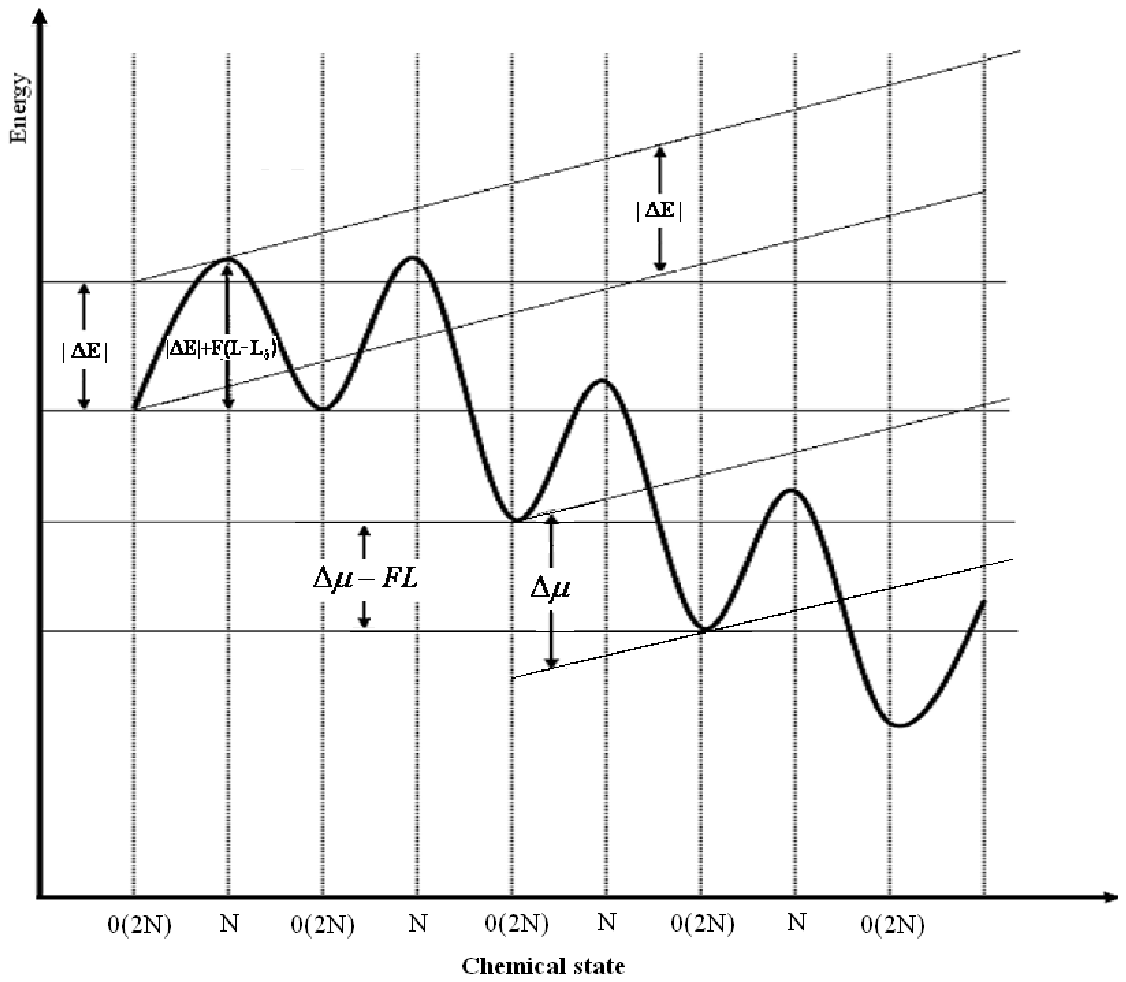}\\
  \caption{
  The energy profile in mechanochemical cycles: {\bf (Left)} No external force case: the energy barrier between mechanochemical states
  $0$ and $N$ is $|\triangle E|$. After mechanochemical state $N$, the molecular motor might back to state $0$ through mechanochemical passway
  $N\to N-1\to\cdots\to 0$ or $N\to N+1\to\cdots\to 2N(0)$. In this case, the molecular motor makes no any mechanical steps. Also, the molecular motor
  might back to state $0$ immediately through mechanochemical passway $N\to 0$. In such case, molecular motor completes one forward mechanical step,
  with one fuel molecule ATP is hydrolyzed. The free energy released by one ATP molecule is $\triangle \mu$.  {\bf (Right)} Nonzero external force $F$ case:
  in this case, the energy barrier between mechanochemical states $0$ and $N$ is $|\triangle E|+F(L-L_\delta)$, which is bigger than the no external force
  case. So it will be more difficult for molecular motors to make a forward step. During one forward step, the energy dissipation is $\triangle \mu-FL$,
  which is small than the no external force case, since part of the energy $\triangle\mu$ released by the ATP is used to do useful mechanical work.}
  \label{Figure4}
\end{figure}

In (\ref{eq37}), the load distribution factors $\delta$ and
$\delta_i$ can be determined by experimental data as in
\cite{Masayoshi2002, Masayoshi2003, Yuichi2005, Zhang2008}.
Thermodynamic consistency requires $k_i^+/k_{i+1}^-=e^{\beta
\triangle E_{i}}$ and $k^+/k^-=e^{\beta (\triangle \mu -\triangle
E)}$, where $\triangle E_{i}=E_i-E_{i+1}$ is the potential energy
difference between mechanochemical states $i$ and $i+1$ (see Figure
\ref{Figure3}), $\triangle E=\sum_{i=0}^{N-1}\triangle E_i=E_0-E_N$
is the potential energy difference between mechanochemical states
$0$ and state $N$, in the no external force case, which is the
energy barrier of the movement of molecular motors. $\triangle \mu$
is the chemical energy transferred to the motors in one
mechachemical step, which comes from the hydrolysis of the fuel
molecule ATP (see Figure \ref{Figure4}).

\section{Continuous mechanochemical state multi-cycle network model}
\begin{figure}
  \includegraphics[width=450pt]{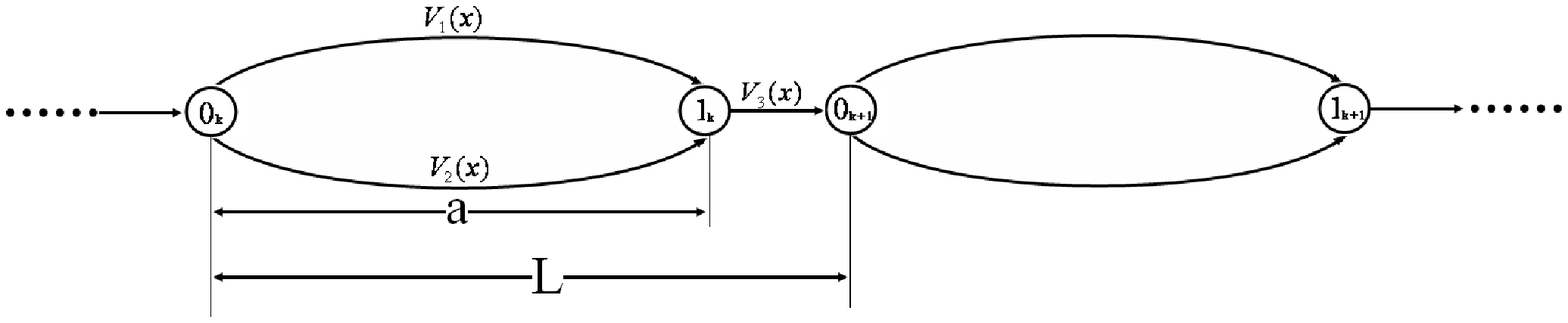}\\
  \caption{Depiction of continuous mechanochemical state multi-cycle network model: There're two chemical passway between mechanochemical state
  $0_k$ and $1_{k}$, in which the potentials are $V_1(x)$ and $V_2(x)$ ($kL\le x\le kL+a$) respectively. The potential between mechanochemical state
  $1_{k}$ and $0_{k+1}$ is $V_3(x)$ ($kL+a\le x\le (k+1)L$).}\label{Figure5}
\end{figure}

As the number of mechanochemical states $2N$ tends to infinite, our
multi-cycle network model (see Figure \ref{Figure1}) can be
approximated by the continuous mechanochemical state model (see
Figure \ref{Figure5}).
In this model, there're two chemical passway from state $0_k$ to
state $1_k$ with different potentials $V_1(x)$ and $V_2(x)$ ($kL\le
x\le kL+a$) respectively. From state $1_k$ to state $0_{k+1}$, the
potential function is $V_3(x)$ ($kL+a\le x\le (k+1)L$).
Biophysically, the potentials $V_i(x)$ are periodical, i.e
$V_i(x+L)=V(x)$, and satisfy $V_1(0_k)=V_2(0_k),
V_1(1_k)=V_2(1_k)=V_3(1_k)$.

In the $i-$th chemical passway, the motion of molecular motors can
be described by the following Fokker-Planck equation:
\begin{equation}\label{eq38}
\begin{aligned}
&\frac{\tilde{\rho}_i(x,t)}{\partial t}=\frac{\partial}{\partial
x}\left(\frac{\tilde{\rho}_i(x,t)}{\gamma}\frac{\partial
V_i(x)}{\partial x}+D\frac{\partial \tilde{\rho}_i(x, t)}{\partial
x}\right)=-\frac{\partial \tilde{J}_i}{\partial x}\cr
&\textrm{where} \qquad kL\le x\le kL+a\qquad\qquad\
\textrm{for}\quad i=1,2\cr &\qquad\qquad \ \ kL+a\le x\le
(k+1)L\qquad\textrm{for}\quad i=3
\end{aligned}
\end{equation}
in which $\gamma$ is viscous friction coefficient, $D$ is free
diffusion coefficient which satisfies Einstein relation
$D=k_BT/\gamma$, $\tilde{\rho}_i(x,t)$ is probability density for
finding molecular motors at mechanochemical state $x$ in $i-$th
passway at time $t$ and $\tilde{J}_i(x,t)$ is the probability flux.
Define
\begin{equation}\label{eq39}
\begin{aligned}
\rho_i(x,t)=\sum_{k=-\infty}^{\infty}\tilde{\rho}_i(x+kL,t)\qquad
J_i(x,t)=\sum_{k=-\infty}^{\infty}\tilde{J}_i(x+kL,t)
\end{aligned}
\end{equation}
it can be readily verified that
\begin{equation}\label{eq40}
\begin{aligned}
&\frac{\rho_i(x,t)}{\partial t}=\frac{\partial}{\partial
x}\left(\frac{\rho_i(x,t)}{\gamma}\frac{\partial V_i(x)}{\partial
x}+D\frac{\partial \rho_i(x, t)}{\partial x}\right)=-\frac{\partial
J_i}{\partial x}\cr &\textrm{where} \qquad 0\le x\le a\qquad
\textrm{for}\quad i=1,2\cr &\qquad\qquad \  a\le x\le
L\qquad\textrm{for}\quad i=3
\end{aligned}
\end{equation}
At steady state, the probability flux $J_i$ is constant and the
probability $\rho_i(x)$ satisfies
\begin{equation}\label{eq41}
\begin{aligned}
&\frac{\partial \rho_i(x, t)}{\partial x}+\frac{\partial
V_i(x)}{\partial x}\frac{\rho_i(x,t)}{k_BT}=-\frac{J_i}{D}\cr
&\textrm{where} \qquad 0\le x\le a\qquad \textrm{for}\quad i=1,2\cr
&\qquad\qquad \ a\le x\le L\qquad\textrm{for}\quad i=3
\end{aligned}
\end{equation}
Under the following constraints
\begin{equation}\label{eq42}
\begin{aligned}
&\int_0^a\rho_1(x)dx+\int_0^a\rho_2(x)dx+\int_a^b\rho_3(x)dx=1\cr
&\rho_1(0)=\rho_2(0)=\rho_3(L)\quad
\rho_1(a)=\rho_2(a)=\rho_3(a)\quad J_1+J_2=J_3
\end{aligned}
\end{equation}
we can get the solutions of (\ref{eq41})
\begin{equation}\label{eq43}
\begin{aligned}
&\rho_i(x)=\left(C_i-\frac{J_i}{D}\int_0^x
e^{\frac{V_i(y)}{k_BT}}dy\right)e^{-\frac{V_i(x)}{k_BT}}\quad
\textrm{for}\ \ i=1,2\cr &\rho_3(x)=\left(C_3-\frac{J_3}{D}\int_a^x
e^{\frac{V_3(y)}{k_BT}}dy\right)e^{-\frac{V_3(x)}{k_BT}}
\end{aligned}
\end{equation}
where the constants $C_1, C_2, C_3$ are the following
$$
\begin{aligned}
&C_1=C_2=\frac{\left[\int_0^a\left(e^{\frac{V_1(y)}{k_BT}}+e^{\frac{V_2(y)}{k_BT}}\right)dy\right]
\left(\int_a^Le^{\frac{V_3(y)}{k_BT}}dy\right)+\left(\int_0^ae^{\frac{V_1(y)}{k_BT}}dy\right)\left(\int_0^ae^{\frac{V_2(y)}{k_BT}}dy\right)}
{ e^{\frac{V_3(L)-V_1(0)}{k_BT}}\Delta}\cr
&C_3=\frac{\left[\int_0^a\left(e^{\frac{V_1(y)}{k_BT}}+e^{\frac{V_2(y)}{k_BT}}\right)dy\right]
\left(\int_a^Le^{\frac{V_3(y)}{k_BT}}dy\right)e^{\frac{V_1(0)-V_3(L)}{k_BT}}+\left(\int_0^ae^{\frac{V_1(y)}{k_BT}}dy\right)\left(\int_0^ae^{\frac{V_2(y)}{k_BT}}dy\right)}
{\Delta}
\end{aligned}
$$
and the probability fluxes $J_1, J_2, J_3$ are
\begin{equation}\label{eq44}
\begin{aligned}
&J_1=\left.\left(e^{\frac{V_1(0)-V_3(L)}{k_BT}}-1\right)\left(\int_0^ae^{\frac{V_2(y)}{k_BT}}dy\right)D\right/\Delta\cr
&J_2=\left.\left(e^{\frac{V_2(0)-V_3(L)}{k_BT}}-1\right)\left(\int_0^ae^{\frac{V_1(y)}{k_BT}}dy\right)D\right/\Delta\cr
&J_3=J_1+J_2
\end{aligned}
\end{equation}
The expression $\Delta$ is
$$
\begin{aligned}
\Delta=&\left[\left(\int_0^ae^{\frac{V_1(y)}{k_BT}}dy\right)\left(\int_0^ae^{\frac{V_2(y)}{k_BT}}dy\right)+
\left(\int_0^ae^{\frac{V_2(y)}{k_BT}}dy\right)\left(\int_a^Le^{\frac{V_3(y)}{k_BT}}dy\right)\right.\cr
&+\left.\left(\int_0^ae^{\frac{V_1(y)}{k_BT}}dy\right)\left(\int_a^Le^{\frac{V_3(y)}{k_BT}}dy\right)\right]\cr
&\times
\left[\int_0^ae^{-\frac{V_1(y)}{k_BT}}dy+\int_0^ae^{-\frac{V_2(y)}{k_BT}}dy+\int_a^Le^{-\frac{V_3(y)}{k_BT}}dy\right]e^{\frac{V_1(0)-V_3(L)}{k_BT}}\cr
&+\left[\left(\int_0^ae^{\frac{V_1(y)}{k_BT}}dy\right)\left(\int_0^ae^{\frac{V_2(y)}{k_BT}}dy\right)\left(\int_a^Le^{-\frac{V_3(y)}{k_BT}}dy\right)\right.\cr
&+\left.\left(\int_0^a\int_0^xe^{\frac{V_2(y)-V_2(x)}{k_BT}}dydx+\int_a^L\int_a^xe^{\frac{V_3(y)-V_3(x)}{k_BT}}dydx\right)\left(\int_0^ae^{\frac{V_1(y)}{k_BT}}dy\right)\right.\cr
&+\left.\left(\int_0^a\int_0^xe^{\frac{V_1(y)-V_1(x)}{k_BT}}dydx+\int_a^L\int_a^xe^{\frac{V_3(y)-V_3(x)}{k_BT}}dydx\right)\left(\int_0^ae^{\frac{V_2(y)}{k_BT}}dy\right)\right]\cr
&\times\left(1-e^{\frac{V_1(0)-V_3(L)}{k_BT}}\right)\cr
=&\left(\int_0^ae^{\frac{V_1(y)}{k_BT}}dy\right)\left(\int_0^ae^{\frac{V_2(y)}{k_BT}}dy\right)\left(\int_a^Le^{-\frac{V_3(y)}{k_BT}}dy\right)\cr
&+\left(\int_0^a\int_0^xe^{\frac{V_2(y)-V_2(x)}{k_BT}}dydx+\int_a^L\int_a^xe^{\frac{V_3(y)-V_3(x)}{k_BT}}dydx\right)\left(\int_0^ae^{\frac{V_1(y)}{k_BT}}dy\right)\cr
&+\left(\int_0^a\int_0^xe^{\frac{V_1(y)-V_1(x)}{k_BT}}dydx+\int_a^L\int_a^xe^{\frac{V_3(y)-V_3(x)}{k_BT}}dydx\right)\left(\int_0^ae^{\frac{V_2(y)}{k_BT}}dy\right)\cr
&+\left\{\left(\int_0^a\int_x^ae^{\frac{V_2(y)-V_2(x)}{k_BT}}dydx+\int_a^L\int_x^Le^{\frac{V_3(y)-V_3(x)}{k_BT}}dydx\right)\left(\int_0^ae^{\frac{V_1(y)}{k_BT}}dy\right)\right.\cr
&+\left(\int_0^a\int_x^ae^{\frac{V_1(y)-V_1(x)}{k_BT}}dydx+\int_a^L\int_x^Le^{\frac{V_3(y)-V_3(x)}{k_BT}}dydx\right)\left(\int_0^ae^{\frac{V_2(y)}{k_BT}}dy\right)\cr
&+\left[\left(\int_0^ae^{\frac{V_2(y)}{k_BT}}dy\right)\left(\int_a^Le^{\frac{V_3(y)}{k_BT}}dy\right)
+\left(\int_0^ae^{\frac{V_1(y)}{k_BT}}dy\right)\left(\int_a^Le^{\frac{V_3(y)}{k_BT}}dy\right)\right]\cr
&\times
\left.\left[\int_0^ae^{-\frac{V_1(y)}{k_BT}}dy+\int_0^ae^{-\frac{V_2(y)}{k_BT}}dy\right]\right\}e^{\frac{V_1(0)-V_3(L)}{k_BT}}\cr
>&0
\end{aligned}
$$
Therefore, in the framework of this continuous mechanochemical state
multi-cycle network model, the expression of the mean velocity of
molecular motors is
\begin{equation}\label{eq45}
V=(J_1+J_2)a+J_3(L-a)=\frac{\left(e^{\frac{V_1(0)-V_3(L)}{k_BT}}-1\right)\left[\int_0^a\left(e^{\frac{V_1(y)}{k_BT}}+e^{\frac{V_2(y)}{k_BT}}\right)dy\right]DL}{\Delta}
\end{equation}
Obviously, $V>0$ if $V_1(0)>V_3(L)$ and $V<0$ if $V_1(0)<V_3(L)$. It
can be readily verified that the equations (\ref{eq1}) (\ref{eq2})
can be obtained by applying spatial discretization to Fokker-Planck
equation (\ref{eq38}), with some detailed expression of the
transition rate $u_i, w_i$ (see \cite{Wang2003}).

\section{Concluding remarks}
In this paper, a general multi-cycle network model of molecular
motors is theoretically discussed. The explicit formulation of the
velocity has been obtained. This model can be regarded as a
generalization of the one designed by Liepelt and Lipowsky in
\cite{Liepelt2009} and the one used by T. Schmiedl and U. Seifert in
\cite{Schmiedl2008}. The method used in this paper is similar as the
methods used by Derrida, Fisher and Kolomeisky \cite{Fisher2001,
Kolomeisky2000,Derrida1983}.

\vskip 0.5cm

\noindent{\bf Acknowledgments}

This work was funded by National Natural Science Foundation of China
(Grant No. 10701029).

%\bibliography{velocity}

\end{document}